\documentclass[10pt,reqno]{amsart}
     \makeatletter
     \def\section{\@startsection{section}{1}%
     \z@{.7\linespacing\@plus\linespacing}{.5\linespacing}%
     {\bfseries
     \centering
     }}
     \def\@secnumfont{\bfseries}
     \makeatother
\newtheorem{theorem}{Theorem}[section]

\newtheorem{proposition}[theorem]{Proposition}

\theoremstyle{definition}

\theoremstyle{remark}

\numberwithin{equation}{section}
\usepackage{amsmath}
\usepackage{amsthm}
\usepackage{amsfonts}
\usepackage{graphicx}
\usepackage{caption}
\usepackage{enumitem}
\usepackage{hyperref}
\hypersetup{
	colorlinks=true,
	linkcolor=blue,
	filecolor=magenta,
	urlcolor=cyan,
	}
\usepackage{float}
\begin{document}

\title[Closed Quantum Black-Scholes]{Closed Quantum Black-Scholes: Quantum Drift and the Heisenberg Equation of Motion.}

\author{Will Hicks}
\address{Will Hicks, Investec Bank PLC, 30 Gresham Street, London EC2V 7QP, United Kingdom}
\email{whicks7940@googlemail.com}




\begin{abstract}
In this article we model a financial derivative price as an observable on a market state function, with a view to understanding how some of the non-commutative behaviour of the financial market impacts the dynamics. We integrate the Heisenberg Equation of Motion, by using a Riemannian metric, and illustrate how the non-commutative nature of the model introduces quantum interference effects that can act as either a drag or a boost on the resulting return. The ultimate objective is to investigate the nature of quantum drift in the Accardi-Boukas quantum Black-Scholes framework (\cite{AccBoukas}) which involves modelling the financial market as a quantum observable, and introduces randomness through the Hudson-Parthasarathy quantum stochastic calculus (\cite{HP}). In particular we aim to differentiate between randomness that is introduced through external noise (quantum stochastic calculus) and randomness that is intrinsic to a quantum system (Heisenberg Equation of Motion).
\end{abstract}

\maketitle

\noindent 

\section{Introduction}
\subsection{The Accardi-Boukas Quantum Black-Scholes:}
The quantum Black-Scholes framework of Accardi \& Boukas (see \cite{AccBoukas}) derives a partial differential equation for the value of a derivative security by applying the quantum stochastic calculus of Hudson \& Parthasarathy (see \cite{HP}). This is a generalisation of the standard Ito stochastic calculus based on modelling the relevant traded underlying as a quantum observable. We note the following:
\begin{itemize}
\item The traded underlying is modelled as an $\mathcal{K}$ valued stochastic process in the Hilbert space: $L^2(\mathbb{R}^+)\otimes\mathcal{K}$.
\item Random noise is added through the Boson Fock space: $\Gamma(L^2(\mathbb{R}^+)\otimes\mathcal{K})$ (we write $\Gamma$).
\item One quantizes the traded asset price as an operator valued observable $X$ on the space: $\mathcal{K}\otimes\Gamma$.
\item Derivative payouts are considered as operator valued functions of the quantum observable.
\end{itemize}
\vspace{3mm}
The benefits of this approach are that one can introduce diffusions within noncommutative spaces which mirror many of the real effects of the financial markets, that are difficult to model using `classical' Ito processes (for example see \cite{Hicks}, \cite{Hicks2}, and \cite{Hicks3}). However, we will see that this particular choice of quantization means that the internal dynamics of the quantum system $\mathcal{K}$ governed by the Hamiltonian $\hat{H}$ are lost, and have no impact on the derivative price.
\vspace{3mm}

The time evolution of the underlying quantum state, that arises from this Hamiltonian, drops out of the analysis in the same way as the classical drift does in the classical Black-Scholes. If the traded underlying price evolves under geometric Brownian motion:
\begin{equation}
dS=\mu Sdt+\sigma SdW^{\mathcal{P}}
\end{equation}
Then $\mu$ is replaced by the funding cost in our chosen Martingale measure: $\mathcal{Q}$. For example, if we choose a deposit account, bearing interest rate: $r$, as our risk free asset, then the risk neutral process used for derivative pricing becomes:
\begin{equation}
dS=r Sdt+\sigma SdW^{\mathcal{Q}}
\end{equation}
Working in the Heisenberg interpretation of Quantum mechanics, the unitary time development operator: $U_t$, which is usually given by:
\begin{equation}
dU_t=\Big(-i\hat{H}dt\Big)U_t
\end{equation}
becomes, after adding random noise through the operators $dA_t, dA^{\dagger}_t, d\Lambda_t$, that fill up the Boson Fock space, combined with bounded linear operators $L$ and $S$, where $S$ is unitary:
\begin{equation}\label{dU}
dU_t=\Big(\big(-i\hat{H}-\frac{1}{2}L^*L\big)dt+L^*SdA_t-LdA^{\dagger}_t+(1-S)d\Lambda_t\Big)U_t
\end{equation}
In the same way that charged particles interact with a quantized electromagnetic field via the creation of photons in a Boson Fock space, the financial market interacts with external environmental noise, through the creation of `noise packets' in a Boson Fock space. The external environment creates `financial photons'.
\vspace{3mm}

The observable: $X$ that acts on the underlying state to return the current price for the traded asset, evolves in time under the unitary operator described by \ref{dU}:
\begin{equation}\label{dX_1}
\begin{split}
j_t(X)=U_t^*XU_t\\
dj_t(X)=dU_t^*XU_t+U_t^*XdU_t+dU_t^*XdU_t
\end{split}
\end{equation}
This in turn can be simplified using a quantum version of Ito's lemma, described by \cite{HP} Theorem 4.5. This result gives a multiplication table that can be used to combine the operators in equation \ref{dX_1}:
\begin{equation}\label{Q_Ito}
\begin{tabular}{p{1cm}|p{1cm}p{1cm}p{1cm}p{1cm}}
-&$dA^{\dagger}_t$&$d\Lambda_t$&$dA_t$&$dt$\\
\hline
$dA^{\dagger}_t$&0&0&0&0\\
$d\Lambda_t$&$dA^{\dagger}_t$&$d\Lambda_t$&0&0\\
$dA_t$&$dt$&$dA_t$&0&0\\
$dt$&0&0&0&0
\end{tabular}
\end{equation}
Using this, we get the following dynamics for $X$:
\begin{equation}\label{dX_2}
\begin{split}
dj_t(X)=j_t(\alpha^{\dagger})dA^{\dagger}_t+j_t(\alpha)dA_t+j_t(\lambda)d\Lambda_t+j_t(\theta)dt\\
\alpha=[L^*,X]S\\
\lambda=S^*XS-X\\
\theta=i[H,X]-\frac{1}{2}(L^*LX+XL^*L-2L^*XL)
\end{split}
\end{equation}
It is important to note the following two things:
\begin{itemize}
\item Under $L=0$, we have a system where no financial photons are being created. There is no environmental noise, and equation \ref{dX_2} simplifies to the Heisenberg equation of motion.
\item Under $L\neq 0$, the Hamiltonian driving the dynamics of the underlying Hilbert space ($\mathcal{K}$), $\hat{H}$, impacts only the drift of the quantum observable.
\end{itemize}
\vspace{3mm}
To derive the quantum Black-Scholes equation, following Accardi \& Boukas (\cite{AccBoukas}), first we expand values for $(dj_t(X))^k$ using \ref{Q_Ito}:
\begin{equation}
(dj_t(X))^k=j_t(\lambda^{k-1}\alpha^{\dagger})dA^{\dagger}_t+j_t(\alpha\lambda^{k-1})dA_t+j_t(\lambda^k)d\Lambda_t+j_t(\alpha\lambda^{k-2}\alpha^{\dagger})dt
\end{equation}
Next, we quantize the derivative valuation, by considering operator valued functions of the quantum observable: $F: [0,T]\times \mathcal{B}(\mathcal{K}\otimes\Gamma)\longrightarrow\mathcal{B}(\mathcal{K}\otimes\Gamma)$. Writing $j_t(X)=x$, we have:
\begin{equation}
F(t,x)=\sum_{n,k\geq 0}\frac{1}{n!k!}\frac{\partial^{n+k} F}{\partial t^n\partial x^k}
\end{equation}
In order to derive the final quantum Black-Scholes, Accardi \& Boukas show how to derive a value process that, by holding units of the risky underlying $j_t(X)$, and a chosen numeraire asset, pays out the final derivative payout with probability 1.
\subsection{Extending the Accardi-Boukas Approach:}
This approach is extremely powerful, since it allows us to extend the existing classical approach to Mathematical Finance to a noncommutative setting. The financial markets are in fact naturally noncommutative:
\begin{itemize}
\item It is largely impossible to determine the price one can trade at, with infinite precision, in advance. For example, when a trader submits a `bid' order to an exchange mechanism, they do not usually know what `offer' prices will be submitted, and therefore what price the `bid' will get filled at.
\item One can only really identify the true price of a traded underlying by submitting a `bid/offer' order, and in so doing, impacting the very state of the market one is trying to measure.
\item As markets are more and more dominated by algorithms, and execution times fall correspondingly, this effect will become more and more pronounced.
\end{itemize}
\vspace{3mm}

The properties of the resulting quantum Black-Scholes equations are further discussed in \cite{Hicks}, \cite{Hicks2}, and \cite{Hicks3}. It has also been noted (for example by Haven in \cite{Haven1}, and discussed further by Melnyk and Tuluzov in \cite{MelTul}) that the risk in trading financial derivatives cannot be fully hedged in a `true' quantum model. When executing the required strategy for the value process, it is not possible to determine the exact price of the traded underlying $j_t(X)$. Furthermore, if our derivative is to be considered a quantum observable in its own right, rather than a function of a quantum observable, then we cannot measure the risk sensitivities: $\partial F/\partial j_t(X)$ with full precision, and therefore cannot evaluate how much of the underlying we need to hold. Moreover, our numeraire asset drifts at a constant rate: $dV=rVdt$. However, constant drift at a pre-specified rate is not something that can be realised in a fully quantum model of the financial market.
\vspace{3mm}

These considerations do not invalidate in any way the resulting quantum Black-Scholes. Whilst in the quantum case, executing the value process that replicates a derivative payout is not possible, this does not prevent the resulting valuation being a unique no arbitrage price for the derivative.
\vspace{3mm}

However, as noted above, the impact of trading activity on the `state' of the market is one of the key reasons why a non-commutative approach makes sense. Therefore, in an ideal world we would like to build a model that is capable of incorporating the impact of measurement on the market state. The impact of trading activity on the evolution of the market state has been discussed previously, for example see \cite{Piotrowski}-\cite{Piotrowski3} where the authors highlight how different market effects can be explained using the non-commutativity of quantum mechanics.
\vspace{3mm}

Furthermore, whilst the quantization chosen by Accardi \& Boukas allows the derivation of the quantum Black-Scholes equation, a natural alternative is to consider derivative payouts as themselves quantum observables, rather than operator valued functions of a quantum observable. Whilst in some respects this distinction is a philosophical choice around the boundary separating the quantum \& classical regimes, we find in this article that it can have a real impact on the model.
\vspace{3mm}

Therefore, the long term goal is to build a model of the financial market based on the following ingredients:
\begin{enumerate}
\item[{\bf A}] A quantum market state, with the dynamics controlled by a Hamiltonian function.
\item[{\bf B}] A set of observables, tradeable instruments, by which one can interact with the market.
\item[{\bf C}] A means by which we can allow the introduction of environmental noise. For example, through the quantum stochastic calculus discussed above.
\end{enumerate}
\vspace{3mm}
In this article we make a start with {\bf A} and {\bf B}, by attempting to develop the idea of quantum drift. In essence, when building a model based on classical stochastic calculus, in the event that no external noise is added (no diffusion term), then the final payout from a derivative security can be determined with full precision. In a quantum framework this is no longer the case. There are in fact 2 sources of randomness. An external source based on quantum stochastic calculus, such as that introduced by Accardi \& Boukas, and purely quantum internal randomness that we investigate here.
\vspace{3mm}

We start in section \ref{CQBS}, by deriving the dynamics for quantized financial observables using the Heisenberg equation of motion. We apply a geometric technique to deriving eigenfunctions, before discussing how recursive techniques (for example see \cite{AFH}) can be used to generate power series solutions to the closed quantum Black-Scholes. This technique involves introducing a Riemannian metric in order to simplify the integration, and functional form for the eigenfunctions.
\vspace{3mm}

We go on to discuss the near classical limit in section \ref{Classical}. In section \ref{Num}, we investigate some numerical examples, in order to understand the behaviour of solutions, and discuss potential applications of the techniques beyond those discussed above. Finally, we draw overall conclusions in section \ref{Conc}.
\section{Closed Quantum Black-Scholes}\label{CQBS}
The approach starts with a Hilbert space: $\mathcal{H}$, that describes the state of the market. For example, we could select $L^2(\mathbb{R})$. In this case, we could define the operator: $X_0$ to measure the expected price for a market transaction that occurs right now ($t=0$). So:
\begin{equation}
X_0\psi(x)=x\psi(x)
\end{equation}
\begin{equation}
E^0[X_0]=\langle\psi|X\psi\rangle=\int_{\mathbb{R}}x|\psi(x)|^2dx
\end{equation}
Following the usual principals of quantum mechanics, in the Heisenberg interpretation, time development of the price observable is controlled by a unitary operator: $U(t)=e^{i\hat{H}t}$, where $\hat{H}$ is the Hamiltonian operator for our system. Now, at time $t\neq 0$, we have:
\begin{equation}
X(t)=e^{i\hat{H}t}X_0e^{-i\hat{H}t}
\end{equation}
\begin{equation}
X(t)\psi(x)=e^{i\hat{H}t}xe^{-i\hat{H}t}\psi(x)
\end{equation}
The expected price for a transaction in the future is now:
\begin{equation}
E^t[X(t)]=\langle\psi|X(t)\psi\rangle=\langle e^{-i\hat{H}t}\psi|X_0e^{-i\hat{H}t}\psi\rangle=\int_{\mathbb{R}}x|\psi(x,t)|^2dx
\end{equation}
Where: $\psi(x,t)=e^{-i\hat{H}t}\psi(x)$. This framework can be applied to any observable on the state of the financial market. For observable $A$, we have:
\begin{equation}\label{obs}
A(t)=e^{i\hat{H}t}Ae^{-i\hat{H}t}
\end{equation}
\begin{equation}
E^t[A(t)]=\langle\psi|A(t)\psi\rangle=\langle e^{-i\hat{H}t}\psi|Ae^{-i\hat{H}t}\psi\rangle=\int_{\mathbb{R}}x|\psi(x,t)|^2dx
\end{equation}
Differentiating equation \ref{obs}, we get the Heisenberg equation of motion for a general observable, (where $[A,B]=AB-BA$):
\begin{equation}\label{Heis}
\frac{dA(t)}{dt}=i[\hat{H},A(t)]
\end{equation}
\subsection{Martingale Pricing}\label{mart_price}
Let $X(t,T)$ represent the market price for a traded asset that is fixed at time $t$, and exchanged at maturity $T$. We consider the observable relating to the price for a derivative contract that pays out an amount $F(X(T,T))$, depending on the final observation of the price: $X(T,T)$.
\vspace{3mm}

We let $U(t,x)$ represent the value for this contract at time $t$, contingent on the current price: $X(t,T)=x$. If $X(t,T)$ can be modelled by some probability space: $(\Omega,\mathcal{P},\mathcal{F})$, we have from the first and second fundamental theorem of mathematical finance (see for example \cite{Bjork}, Theorems 10.14, 10.17, 10.18):
\vspace{3mm}

\begin{itemize}
\item The market is arbitrage free if and only if there exists a Martingale measure: $\mathcal{Q}\sim \mathcal{P}$.
\item The market is complete, so all derivative contracts $U(t,x)$ can be replicated by trading in $X(t,T)$, if and only if the Martingale measure is unique. This is the case, assuming the number of underlying random variables that drive market prices, is the same as the number of traded assets. In this simple case: $1$.
\item The value of the derivative contract is given by the discounted expectation in the Martingale measure.
\end{itemize}
\vspace{3mm}
Therefore, if we have an interest rate: $r$, and discount factors: $e^{-r(T-t)}$, we have:
\begin{equation}\label{mart}
U(t,x)=e^{-r(T-t)}E^{\mathcal{Q}}[F(X(T,T))]
\end{equation}
We now turn back to equation \ref{Heis}, and consider the requirements for this equation to represent a pricing equation for a quantum observable for the derivative payout: $F$. We write the observable $X(t,T)=X$, and insert into equation \ref{Heis}:
\begin{equation}
\frac{dX}{dt}=i[\hat{H},X]
\end{equation}
We assume our Hamiltonian operator incorporates a potential energy component, $V$, which acts by pointwise multiplication by $x$. In other words:
\begin{equation}\label{Ham}
\hat{H}\psi(x)=-\frac{\sigma^2}{2}\frac{\partial^2\psi}{\partial x^2}+V(x)\psi(x)
\end{equation}
So:
\begin{equation}
i[\hat{H},X]\psi(x)=-i\sigma^2\frac{\partial\psi}{\partial x}
\end{equation}
If we define a `momentum' operator: $\big(\hat{P}\psi\big)(x)=-i\frac{\partial\psi}{\partial x}$, we get:
\begin{equation}
\frac{dX}{dt}=\sigma^2\hat{P}
\end{equation}
Therefore, {\em classically speaking}, the relevant Martingale measure is that which drifts with the market price. I.e, the measure under which the momentum is zero. From a quantum perspective we have:
\begin{equation}
\frac{d\hat{P}}{dt}=i[\hat{H},\hat{P}]=-\frac{dV}{dX}
\end{equation}
So, in the absence of any potential function, the expected rate of change in the momentum operator is zero. In our quantum framework, we cannot insist the momentum is zero. However, if we take as an initial condition that the expected momentum is zero, then we will have $E^0[X]=E^T[X]$, and our measure will be a Martingale measure. Thus, even though we cannot apply the delta hedging argument used in the original derivation of the Black-Scholes partial differential equation (see \cite{BS},\cite{Merton}), we can still construct a Martingale measure, and consequently an arbitrage free price.\vspace{3mm}

In other words, whilst we cannot know, with full precision, the nature of the self financing almost-simple strategy with which we can replicate a payout, the fundamental theorem of mathematical finance (for example see \cite{Bjork} Theorem 10.5) still guarantees that we cannot execute a self financing strategy $h(t)$, such that the final payout $U(h(T))$ satisfies the following:
\begin{itemize}
\item $P\big(U(h(T))\geq 0\big)=1$
\item $P\big(U(h(T))>0\big) >0$.
\end{itemize}
\subsection{Heisenberg Equation of Motion}
We can now apply the Heisenberg approach to derive a closed quantum Black-Scholes equation for the derivative price. Let $U$ represent the price for our derivative contract. In order to derive the required equation, we must first construct a suitable Hilbert space operator from this classical function.
\vspace{3mm}

Starting from the standard position operator $X$, we need to invoke the spectral theorem (see for example \cite{Hall} Theorem 7.12) in order to represent $U$ as a function of this position operator: $U(X,t)$. In fact, complications arise from the fact that both $X$, and $U(X,t)$ are unbounded. For the purposes of this article, we ignore these complications, and proceed on a formal basis.
\vspace{3mm}

Thus we assume we can form an operator $U(X,t)$, that acts on the market state by pointwise multiplication by an unknown function of $x$ and $t$. We then have:
\begin{equation}
E^{\psi}[U]=\langle\psi(x),e^{it\hat{H}}Ue^{-it\hat{H}}\psi(x)\rangle=\langle e^{-it\hat{H}}\psi(x),Ue^{-it\hat{H}}\psi(x)\rangle
\end{equation}
So, applying the Heisenberg equation of motion:
\begin{equation}
E^{\psi}\bigg[\frac{\partial U}{\partial t}\bigg]=\langle e^{-it\hat{H}}\psi(x),i[\hat{H},U]e^{-it\hat{H}}\psi(x)\rangle
\end{equation}
If we assume the potential energy component from \ref{Ham}, commutes with the derivative price operator, $[V,U]=0$, then we have:
\begin{equation}
\begin{split}
i[\hat{H},U]=-i\sigma^2\frac{\partial U}{\partial x}\frac{\partial}{\partial x}-\frac{i\sigma^2}{2}\frac{\partial^2 U}{\partial x^2}\\
=\sigma^2\frac{\partial U}{\partial x}\hat{P}+\frac{\sigma^2}{2}\hat{P}\frac{\partial U}{\partial x}
\end{split}
\end{equation}
So our closed Quantum Black-Scholes becomes:
\begin{proposition}\label{HEM}
Let $X(t,T)$ represent the forward price, observed at $t$ for the purchase of an asset at $T$, and let $U(x,t)$ represent the price for the derivative payout, at time $t$, contingent on the current forward price: $x=X(0,T)$. Then we have:
\begin{equation}\label{pde}
E^{\psi}\Big[\frac{\partial U}{\partial t}\Big]=E^{\psi}\Big[\frac{\sigma^2}{2}\Big(\hat{P}\frac{\partial U}{\partial x}+\frac{\partial U}{\partial x}\hat{P}\Big)\Big]
\end{equation}
Where, for operator $A$:
\begin{equation}
E^{\psi}[A]=\langle \psi,A\psi\rangle
\end{equation}
\vspace{3mm}
Equation \ref{pde}, will yield an arbitrage free price if and only if we have: $E^{\psi}[\hat{P}]=0$, and $[\hat{H},\hat{P}]=0$.
\end{proposition}
Note the following:
\begin{itemize}
\item The derivative price operator: $U$ is now dependent on the state function. Now, rather than making assumptions about the stochastic process that any traded price follows, the price can be derived by making assumptions about the market state, and associated Hamiltonian dynamics.
\item If we choose $\mathcal{H}=L^2(\mathbb{R})$, it is possible to show for reasonable cases, that the operator $\hat{P}\frac{\partial U}{\partial x}+\frac{\partial U}{\partial x}\hat{P}$ is self-adjoint on a dense domain (see \cite{GP}, \cite{BoukasFein}). This enables us to calculate associated eigenfunctions that can be used to integrate numerically.
\item The potential energy component from the Hamiltonian (\ref{Ham}) controls the {\it classical} drift of the system. By assuming that this commutes with $U$ we ensure that the potential does not impact the no-arbitrage price of the derivative.
\end{itemize}
\subsection{Integrating The Heisenberg Equation of Motion}
Expanding out the right hand side of equation \ref{pde}, we get:
\begin{equation}
\Big\langle\psi,\frac{\partial U}{\partial t}\psi\Big\rangle=\Big\langle \psi,-i\frac{\sigma^2}{2}\frac{\partial^2 U}{\partial x^2}\psi\Big\rangle-\Big\langle\psi, i\sigma^2\frac{\partial U}{\partial x}\frac{\partial \psi}{\partial x}\Big\rangle
\end{equation}
\subsubsection{Naive Approach:}
To find eigenfunctions, we must solve the equation (for $\psi$):
\begin{equation}
-i\sigma^2\frac{\partial U}{\partial x}\frac{\partial\psi}{\partial x}-\frac{i\sigma^2}{2}\frac{\partial^2 U}{\partial x^2}\psi-\lambda\psi=0
\end{equation}
\vspace{3mm}
Rearranging, we have:
\begin{equation}
\frac{1}{\psi}\frac{\partial\psi}{\partial x}=\frac{2i\frac{\lambda}{\sigma^2}-\frac{\partial^2 U}{\partial x^2}}{2\frac{\partial U}{\partial x}}
\end{equation}
Applying the integrating factor method, we can calculate the eigenfunctions:
\begin{equation}
\phi_{\lambda}(x)=\frac{1}{\sqrt{\partial U/\partial x}}exp\bigg(i\frac{\lambda}{\sigma^2}\int_0^x(\partial U/\partial x(s))^{-1}ds\bigg)
\end{equation}
The functional form for this solution is suggestive of using a new distance metric: $g(x)=\partial U/\partial x$ to solve the problem. 
We investigate this approach in the next section.
\subsubsection{Geometric Approach:}\label{Cases}
The analysis above, suggests using a distance metric: $g(x)$ to solve \ref{HEM}. The general form for a Laplacian operator on a 1 dimensional Riemannian manifold is given by (see for example \cite{PHL} chapter 4):
\begin{equation}\label{general_LP}
\Delta_g=g^{-1/2}\Big(\frac{\partial}{\partial x}+\mathcal{A}_x\Big)g^{-1/2}\Big(\frac{\partial}{\partial x}+\mathcal{A}_x\Big)+Q(x)
\end{equation}
where $\mathcal{A}_x$ represents the components of an Abelian connection, and $Q(x)$ a section of a real vector bundle over our manifold. If we select $\mathcal{A}_x=Q(x)=0$, we end up with the Laplace-Beltrami operator:
\begin{equation}\label{Laplace_Belt}
\frac{\sigma^2}{2\sqrt{g(x)}}\frac{\partial}{\partial x}\bigg(\frac{1}{\sqrt{g(x)}}\frac{\partial}{\partial x}\Big(...\Big)\bigg)
\end{equation}
However, our objective is to use geometry to simplify the problem, to enable us to derive eigenfunctions:
\begin{itemize}
\item By introducing a distance metric: $g(x)$ we are translating simple Euclidean space into a Riemannian manifold. The Euclidean notion of distance is stretched by different amounts at different points.
\item By choosing $\mathcal{A}_x,Q(x)$, we are choosing a convenient coordinate system on our manifold. Choosing nonzero $\mathcal{A}_x,Q(x)$ is the equivalent for the Riemannian manifold, to choosing curvilinear coordinates in Euclidean space.
\end{itemize}
\vspace{3mm}

Expanding out the general Laplacian, \ref{general_LP}, we get:
\begin{equation}
\frac{1}{g(x)}\frac{\partial^2}{\partial x^2}+\bigg(g(x)^{-1/2}\frac{\partial (g(x)^{-1/2})}{\partial x}+\frac{\mathcal{A}_x}{g(x)}\bigg)\frac{\partial}{\partial x}+\bigg(\frac{\mathcal{A}_x^2}{g(x)}+\frac{1}{g(x)}\frac{\partial (\mathcal{A}_x)}{\partial x}+Q(x)\bigg)
\end{equation}
Therefore by choosing:
\begin{equation}
\begin{split}
\mathcal{A}_x=-g^{1/2}\frac{\partial (g(x)^{-1/2})}{\partial x}\\
Q(x)=-\frac{\mathcal{A}_x^2}{g(x)}-\frac{1}{g(x)}\frac{\partial (\mathcal{A}_x)}{\partial x}
\end{split}
\end{equation}
The operator is simplified to:
\begin{equation}
\Delta_g=\frac{1}{g(x)}\frac{\partial^2}{\partial x^2}
\end{equation}
Under this coordinate system our Hamiltonian becomes (we use a slight abuse of notation, by still writing the independent variable as $x$):
\begin{equation}
\hat{H}\psi(x)=-\frac{\sigma^2}{2g(x)}\frac{\partial^2\psi}{\partial x^2}
\end{equation}
Equation \ref{pde} now becomes:
\begin{equation}
\Big\langle\psi,\frac{\partial U}{\partial t}\psi\Big\rangle=\Big\langle \psi,-i\frac{\sigma^2}{2g(x)}\frac{\partial^2 U}{\partial x^2}\psi\Big\rangle-\Big\langle\psi, \frac{i\sigma^2}{g(x)}\frac{\partial U}{\partial x}\frac{\partial \psi}{\partial x}\Big\rangle
\end{equation}
Setting $g(x)=\partial U/\partial x$, we get:
\begin{equation}
\bigg(\frac{-i\sigma^2}{2g}\frac{\partial g}{\partial x}-\lambda\bigg)=\frac{i\sigma^2}{\psi}\frac{\partial\psi}{\partial x}
\end{equation}
Multiplying by $-i/\sigma^2$, and integrating we get:
\begin{equation}\label{eig_fun}
\sqrt{g(x)}\psi(x)=e^{i\lambda x/\sigma^2}
\end{equation}

\subsection{Solutions to the Closed Quantum Black-Scholes}
We write the Heisenberg equation of motion as:
\begin{equation}\label{Informal_HEM}
\frac{\partial U}{\partial t}=i[\hat{H},U]
\end{equation}
A formal solution to this equation can be written (see \cite{AFH}):
\begin{equation}\label{FullSoln}
\begin{split}
U(t)=exp(iLt)U(0)\\
LU=[\hat{H},U]
\end{split}
\end{equation}
Power series solutions to this equation are discussed below. However, we can use the eigenfunctions to solve for the short time behaviour for $U(t)$. The first step is to expand the market state over the eigenfunctions (\ref{eig_fun}):
\begin{equation}\label{FT}
\sqrt{g(x)}\psi(x)=\frac{1}{\sigma^2}\int_{\mathbb{R}}\tilde{\psi}(\lambda/\sigma^2)e^{i\lambda x/\sigma^2}d\lambda
\end{equation}
Applying the spectral theorem at a purely formal level (for example, see \cite{Hall} Theorem 10.9, 10.10) we get:
\begin{equation}\label{U_0_1}
\begin{split}
E^{\psi}\big[U(t)\big]\approx\int_{\mathbb{R}}U_0(x)|\psi(x)|^2dx+\frac{1}{\sigma^2}\Big(\int_{\mathbb{R}}\lambda |\tilde{\psi}(\lambda/\sigma^2)|^2d\lambda\Big)t+O(t^2)
\end{split}
\end{equation}
Strictly speaking, $g(x)$ will vary with time, as the function $U=U(x,t)$ varies. However, \ref{FullSoln} represents an expansion about the initial value: $g(x,0)$.
\subsubsection{Phase Factor Choice, and Quantum Interference}
One of the requirements for the model to be arbitrage free, is that the expected momentum is zero. Therefore we require:
\begin{equation}\label{momentum}
\begin{split}
\int_{\mathbb{R}}k|\tilde{\psi}(k)|^2dk=0\\
\tilde{\psi}(k)=\int_{\mathbb{R}}\psi(x)e^{-ikx}dx
\end{split}
\end{equation}
Having a real valued market state function: $\psi(x)$ will ensure this is the case. In this event, equation \ref{U_0_1} will result in a distribution: $|\tilde{\psi}(\lambda/\sigma^2)|^2$ that is an even function of $\lambda/\sigma^2$. Therefore we end up with:
\begin{equation}
\begin{split}
\frac{1}{\sigma^2}\Big(\int_{\mathbb{R}}\lambda |\tilde{\psi}(\lambda/\sigma^2)|^2d\lambda\Big)=0\\
E^{\psi}\big[U(t)\big]\approx\int_{\mathbb{R}}U_0(x)|\psi(x)|^2dx+O(t^2)
\end{split}
\end{equation}
So, modulo the higher order terms, whilst the quantum framework introduces unhedgeable uncertainty, the expected return on the derivative is not impacted. However, we can meet the condition \ref{momentum} by using a real function multiplied by a phase factor: $exp(i\phi(x))$ where $\phi(x)$ is an even function of $x$. For example (with normalising constant $C$, and real $\alpha$): $\psi(x)=Cexp((i\alpha-1)x^2/2\sigma^2)$. In section \ref{Num}, we will see that this introduces a non-zero expected return on the derivative that has no classical counterpart.
\begin{equation}
\frac{1}{\sigma^2}\Big(\int_{\mathbb{R}}\lambda |\tilde{\psi}(\lambda/\sigma^2)|^2d\lambda\Big)\neq 0
\end{equation}
In these instances, there is a non-zero expected return on the derivative, even where the underlying is Martingale and there is no source of external noise. This return can be positive where the quantum interference boosts the derivative valuation, or it can be negative, whereby the non-commutativity yields a holding cost associated with the derivative.
\subsubsection{Brief Comment on the Topology of the Transformation}
It is clear from above, that where the option delta: $\partial U/\partial x=0$, the Laplacian used is not defined. The transformation squashes all points on the real number line representation of the asset price where the option delta is zero, to a single point. Furthermore, where $\partial U/\partial x=0$ is negative, the direction of the distance metric also reverses. So for example:
\begin{itemize}
\item For a call option payout at final maturity ($t=T$ say), the full real number line is squashed to the positive real number line, with all points below the strike to zero.
\item For a straddle payout, where the delta is negative below the strike and positive above the strike, the real number line is bent back in on itself.
\end{itemize}
\vspace{3mm}

For these reasons, we must split any payout, or hedging strategy, into the following:
\begin{itemize}
\item Prices along the real number line, where we are a buyer ($\partial U/\partial x>0$).
\item Prices along the real number line, where we have no interest in buying or selling ($\partial U/\partial x=0$).
\item Prices along the real number line, where we are a seller ($\partial U/\partial x<0$).
\end{itemize}
\vspace{3mm}

For European payouts, this is simple. One can construct any payout as a linear sum of observables with monotonic final delta values.
\vspace{3mm}

For path dependent options, such as American options or barrier options, the situation is more complicated. An American straddle can no longer be written as the linear sum of an American put, and an American call. In this case a degree of entanglement is required. For example, the payout of an American put is contingent on the value of the American call, and vice versa. For a European straddle, both options exist independently from the other. For the American straddle, the existence of one, is contingent on the state of the other. We defer further discussion of this topic to a future article.
\subsubsection{Building a Power Series Solution to the Closed Quantum Black-Scholes.} 
Following the analysis in \cite{AFH}, we set $L$ as the Liouvillian superoperator, so that for general operator $Q(t)$ we have: $LQ(t)=[\hat{H},Q(t)]$.
\vspace{3mm}

Then, conditional on convergence, the solution to \ref{Informal_HEM} can be written:
\begin{equation}\label{expansion}
\begin{split}
U(t)=\sum_{k\geq 0}\frac{(it)^k}{k!}U_k\\
U_k=L^kU(0)
\end{split}
\end{equation}
In this way, we can broadly proceed as follows:
\begin{itemize}
\item Set $g(x)$ in \ref{FT} based on the final option payout, and carry out a change in variables from $t$ to $\tau=T-t$, for final maturity $T$.
\item Decide on the market probability density function, and associated state function: $\psi(x)$.
\item Calculate $\tilde{\psi}(\lambda/\sigma^2)$ from the inverse Fourier transform of \ref{FT}.
\end{itemize}
\vspace{3mm}
So for example, set $U(x,0)=(x-K)^+$ for a vanilla call option, and further set $\tau=T-t$ for final maturity $T$. From equation \ref{FT}, we get:
\begin{equation}
\begin{split}
\tilde{\psi}(\lambda/\sigma^2)=\int_{\mathbb{R}}1_{x>K}\psi(x)e^{-i\lambda x/\sigma^2}dx\\
=\int_{K}^{\infty}\psi(x)e^{-i\lambda x/\sigma^2}dx
\end{split}
\end{equation}
This value for $\tilde{\psi}(\lambda/\sigma^2)$ can then be inserted into \ref{U_0_1}, to generate a first order call price expansion, valid for small time to maturity.
\subsection{Calculating the Higher Order Terms.}
To calculate the higher order terms in the expansion, one would have to work out eigenfunctions of the operator: $L^kU$. For each consecutive term in the expansion \ref{expansion}, we must expand the market state function over a new eigenfunction, and use the resulting transform (denoted $\tilde{\psi}_k(\lambda)$ say) to calculate $U_k$. We can apply the same geometric technique for each of the terms in the recursive solution.
\vspace{3mm}

In general, to obtain the eigenfunctions which we will use to get the $k$th order term, we expect to end up with a differential equation to solve as follows:
\begin{equation}\label{k_pde}
\sum_{j=0}^{k}p_{j}(x)\frac{\partial^j \psi}{\partial x^j}-\lambda\psi=0
\end{equation}
Where the functions $p_j(x)$ derive from the payout in question: $U$, and its derivatives. Writing PDE: \ref{k_pde} in standard form (for example see \cite{Bender} chapter 3) we get:
\begin{equation}\label{k_pde_2}
\frac{\partial^{k} \psi}{\partial x^{k}}+\sum_{j=0}^{k-1}\frac{p_{j}(x)}{p_{k}(x)}\frac{\partial^j \psi}{\partial x^j}-\frac{\lambda}{p_{k}(x)}\psi=0
\end{equation}
Equation \ref{k_pde_2} is likely singular, in the event that there are zeros in the function $p_{k}(x)$. Assuming $U$ and its derivatives are not polynomials of finite order, the equation is also irregular, since it will remain singular after multiplying by $x^N$ for all $N$.
\vspace{3mm}

We can resolve this issue, by using a judicious choice of Riemannian metric, turning $p_{k}(x)$ into a constant. After we have done this, if we choose a payout $U$ with continuous derivatives that do not blow up, then we will be able to solve the resulting equation, and find appropriate eigenfunctions for the operator: $L^kU$.
\vspace{3mm}

It should be noted that for many payouts that are traded (including vanilla call options, European digital options etc), there are discontinuous derivatives, and so strictly further work is still required. Noting that caution is required, we move on and show how to calculate the second order term: $L^2U$, for illustration purposes.\vspace{3mm}

We have: $L^2U\psi=[\hat{H},[\hat{H},U]]\psi$. We write the Hamiltonian as: $\hat{H}=-\frac{\sigma^2}{2}D$, for an elliptic operator $D$ yet to be determined. For now, we write $\hat{H}=a_2(x)\partial^2/\partial x^2+a_1(x)\partial/\partial x+a_0(x)$ and end up with:
\begin{equation}
[\hat{H},U]=a_2(x)\frac{\partial^2U}{\partial x^2}\psi+2a_2(x)\frac{\partial U}{\partial x}\frac{\partial\psi}{\partial x}+a_1(x)\frac{\partial U}{\partial x}
\end{equation}
And, we have:
\begin{equation}\label{long}
\begin{split}
[\hat{H},[\hat{H},U]]=a_2(x)\frac{\partial^2}{\partial x^2}\Big(a_2(x)\frac{\partial^2U}{\partial x^2}\psi+2a_2(x)\frac{\partial U}{\partial x}\frac{\partial\psi}{\partial x}+a_1(x)\frac{\partial U}{\partial x}\psi\Big)\\
+a_1\frac{\partial}{\partial x}\Big(a_2(x)\frac{\partial^2U}{\partial x^2}\psi+2a_2(x)\frac{\partial U}{\partial x}\frac{\partial\psi}{\partial x}+a_1(x)\frac{\partial U}{\partial x}\psi\Big)\\
+a_0(x)\Big(a_2(x)\frac{\partial^2U}{\partial x^2}\psi+2a_2(x)\frac{\partial U}{\partial x}\frac{\partial\psi}{\partial x}+a_1(x)\frac{\partial U}{\partial x}\psi\Big)\\
-a_2(x)\frac{\partial^2U}{\partial x^2}\Big(a_2(x)\frac{\partial^2\psi}{\partial x^2}+a_1(x)\frac{\partial\psi}{\partial x}+a_0(x)\psi\Big)\\
-2a_2(x)\frac{\partial U}{\partial x}\frac{\partial}{\partial x}\Big(a_2(x)\frac{\partial^2\psi}{\partial x^2}+a_1(x)\frac{\partial\psi}{\partial x}+a_0(x)\psi\Big)\\
-a_1(x)\frac{\partial U}{\partial x}\Big(a_2(x)\frac{\partial^2\psi}{\partial x^2}+a_1(x)\frac{\partial\psi}{\partial x}+a_0(x)\psi\Big)
\end{split}
\end{equation}
From equation \ref{long} we can collect together terms in: $\partial^k\psi/\partial x^k$. For $\partial^3\psi/\partial x^3$, we get:
\begin{equation}
\Big(2a_2(x)^2\frac{\partial U}{\partial x}-2a_2(x)^2\frac{\partial U}{\partial x}\Big)=0
\end{equation}
So the coefficient of $\partial^3\psi/\partial x^3$ is zero.
\vspace{3mm}

The coefficient for $\partial^2\psi/\partial x^2$ is given by:
\begin{equation}\label{k_2}
4a_2(x)^2\frac{\partial^2U}{\partial x^2}+2a_2(x)\frac{\partial a_2(x)}{\partial x}\frac{\partial U}{\partial x}
\end{equation}
This can be rearranged to get:
\begin{equation}\label{k_2_2}
4a_2(x)\frac{\partial}{\partial x}\Big(a_2(x)\frac{\partial U}{\partial x}\Big)-2a_2(x)\Big(\frac{\partial a_2}{\partial x}\frac{\partial U}{\partial x}\Big)
\end{equation}
If we assume, that our Laplacian operator takes the form: $a_2(x)\frac{\partial^2}{\partial x^2}$, as is the case in section \ref{Cases}, then we have $a_1(x)=a_0(x)=0$. If this is the case, the coefficient for $\partial\psi/\partial x$ is given by:
\begin{equation}\label{k_1}
2a_2(x)\frac{\partial}{\partial x}\Big(a_2\frac{\partial^2 U}{\partial x^2}\Big)+2a_2(x)\frac{\partial^2}{\partial x^2}\Big(a_2\frac{\partial U}{\partial x}\Big)
\end{equation}
Now, if we insert the Riemannian metric from above, so that:
\begin{equation}
a_2(x)=-\frac{\sigma^2}{2}\bigg(\frac{\partial U}{\partial x}\bigg)^{-1}
\end{equation}
We find \ref{k_1} can be written:
\begin{equation}
\begin{split}
2a_2(x)\frac{\partial}{\partial x}\Big(a_2\frac{\partial^2 U}{\partial x^2}\Big)\\
=\frac{\sigma^4}{2}\Big(\frac{\partial U}{\partial x}\Big)^{-1}\frac{\partial}{\partial x}\bigg(\Big(\frac{\partial U}{\partial x}\Big)^{-1}\frac{\partial^2 U}{\partial x^2}\bigg)
\end{split}
\end{equation}
Similarly, \ref{k_2_2} becomes:
\begin{equation}
\begin{split}
-2a_2(x)\Big(\frac{\partial a_2}{\partial x}\frac{\partial U}{\partial x}\Big)\\
=\frac{\sigma^4}{2}\Big(\frac{\partial U}{\partial x}\Big)^{-1}\bigg(\Big(\frac{\partial U}{\partial x}\Big)^{-1}\frac{\partial^2 U}{\partial x^2}\bigg)
\end{split}
\end{equation}
Finally, the zero order term in \ref{long}, with $a_0=a_1=0$, is given by:
\begin{equation}
\frac{\sigma^4}{4}\Big(\frac{\partial U}{\partial x}\Big)^{-1}\frac{\partial^2}{\partial x^2}\bigg(\Big(\frac{\partial U}{\partial x}\Big)^{-1}\frac{\partial^2 U}{\partial x^2}\bigg)
\end{equation}
Pulling this all together, we find the 2nd order eigenvalue equation for $L^2U$ is a Sturm-Liouville problem (for example see \cite{Dettman} chapter 7.7).
\vspace{3mm}

$(L^2U)\psi-\lambda\psi=0$ becomes:
\begin{equation}\label{SL}
\begin{split}
\frac{\partial}{\partial x}\bigg(p(x)\frac{\partial\psi}{\partial x}\bigg)+\Big(\frac{\partial^2p}{\partial x^2}\Big)\psi-\Big(\frac{\partial U}{\partial x}\Big)\frac{\lambda}{\sigma^4}\psi=0\\
p(x)=\frac{\partial^2 U/\partial x^2}{\partial U/\partial x}
\end{split}
\end{equation}
Finally, following \cite{Dettman}, we can write this equation in a more amenable shape by using the transformation: $\phi(x)=h(x)\psi(x)$, $h(x)=\big(p(x)\partial U/\partial x\big)^{1/4}$, and the coordinate function: $s=\int_a^x\big(\partial U/\partial x\big)\big(\partial^2 U/\partial x^2\big)^{-1/2}dx$, for arbitrary $a$. Under this transformation \ref{SL} can be written:
\begin{equation}\label{WKB}
\begin{split}
\frac{\partial^2 \phi}{\partial s^2}+\Big(\frac{\lambda}{\sigma^4}-Q(s)\Big)\phi=0\\
Q(s)=\frac{1}{h}\frac{\partial^2 h}{\partial s^2}+\frac{\partial^2 p}{\partial x^2}\frac{1}{\partial U/\partial x}
\end{split}
\end{equation}
Equation \ref{WKB} can then be tackled using the WKB approximation methods outlined in \cite{Hall} chapter 15. We defer this detailed analysis to a future work.
\section{Impact of Quantum Interference \& Discussion of the Classical Limit}\label{Classical}
\subsection{Near Classical Limit}
Equation \ref{U_0_1} represents the spread of returns $\lambda$ that result from purely internal quantum effects. In a Martingale model, the classical limit, given by: $\lim_{\sigma\rightarrow\infty}$, results in the following:
\begin{equation}\label{class}
\begin{split}
E^{\psi}\big[U(t)\big]\approx\int_{\mathbb{R}}U_0(x)|\psi(x)|^2dx+\lim_{\sigma\rightarrow\infty}\bigg(\frac{1}{\sigma^2}\Big(\int_{\mathbb{R}}\lambda |\tilde{\psi}(\lambda/\sigma^2)|^2d\lambda\Big)t\bigg)\\
=\int_{\mathbb{R}}U_0(x)\delta(x_0-x)^2dx\\
=U_0(x_0)
\end{split}
\end{equation}
So we have that the classical limit represents a model with zero randomness (internal or external) and so the expected payout is given by the payout at the initial value. In other words, the classical system isolated from external noise results in zero uncertainty in the final payout. The quantum system isolated from external noise, has a degree of (positive or negative) time value given by a purely quantum interference.\vspace{3mm}

In the near classical limit, where $\sigma^2$ is large in comparison to the variance embedded in the market state:
\begin{equation}\label{ClassImpact}
\sigma^2>>\int_{\mathbb{R}}x^2|\psi(x)|^2dx
\end{equation}
We note the following:
\begin{itemize}
\item The presence of $\sigma^2$ in the Fourier transform term $\tilde{\psi}(\lambda/\sigma^2)$ will ensure the expected return is more and more localised.
\item We intuitively expect the average quantum rate of return:
\vspace{3mm}

$\mu_{\lambda}=\frac{1}{\sigma^2}\Big(\int_{\mathbb{R}}\lambda |\tilde{\psi}(\lambda/\sigma^2)|^2d\lambda\Big)$
\vspace{3mm}

to be $O(1/\sigma^2)$. This would imply that in the limit of large $\sigma^2$ (relative to the variance of the state function), the quantum term will become increasingly localised around zero.
\item The $k^{th}$ order term in the expansion \ref{expansion}, will result in a term in eigenvalues of order: $\frac{\lambda}{\sigma^{2k}}$, originating from $k$ applications of the Hamiltonian function. This implies that in the near clasical limit of large $\sigma$, we can ignore higher order terms.
\item If we write: $\sigma_{\psi}^2=\int_{\mathbb{R}}x^2|\psi(x)|^2dx$, then the ratio $\sigma_{\hat{H}}/\sigma_{\psi}$, where $\sigma_{\hat{H}}$ is the volatility parameter in the Hamiltonian, represents a scale parameter that determines the importance of the quantum interference terms. Where $\sigma_{\hat{H}}/\sigma_{\psi}$ is small, it is more likely the quantum interference will be less important, relative to the external sources of noise.
\item In reality, the length of time between observations of the system (executed trades) is likely to be small in most cases. This is further reason to ignore the higher order terms in the expansion \ref{expansion}.
\end{itemize}
We find overall that a localised market state function, or a low volatility Hamiltonian function, results in a model that is more classical in the sense that the internal quantum interference has less impact. A less localised market state, and high volatility Hamiltonian function results in a model that is more quantum, in the sense that the internal quantum interference has more impact.
\section{Numerical Examples}\label{Num}
In this section we use three numerical examples to illustrate the points made above.
\subsection{Real Valued Wave Function:}
Where the wave function is real valued, the expected return distribution; $\tilde{\psi}(\lambda/\sigma_{\hat{H}}^2)$ will be an even function. Therefore, although there will be quantum interference in the time development of the derivative price observable: $U$, the expected impact is zero. In the chart below, we set $\psi(x)=(1/\sqrt{2\pi\sigma_S^2})exp(-x^2/2\sigma_S^2)$. In this case, $\sigma_S=\sigma_{\hat{H}}=0.2$. We find that setting $g(x)=\partial U/\partial x=1_{x>0}$ increases the uncertainty in the quantum interference. However, the expected interference is still zero.
\begin{figure}[H]
\includegraphics[scale=0.7]{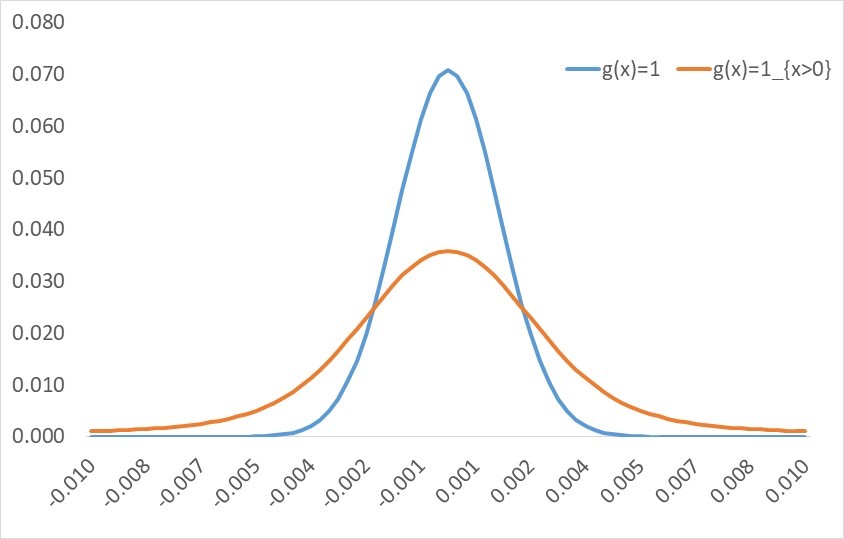}
\caption{The chart shows $\tilde{\psi}(\lambda/\sigma_{\hat{H}}^2)$ for $g(x)=1$, and $g(x)=1_{x>0}$. In both cases we have set the wave function to $\psi(x)=(1/\sqrt{2\pi\sigma_S^2})exp(-x^2/2\sigma_S^2)$.}
\end{figure}
\subsection{Non-Trivial Phase Factor:}
We now apply a phase factor to the wave function: $\psi(x)=(1/\sqrt{2\pi\sigma_S^2})exp((i-1)x^2/2\sqrt{2}\sigma_S^2)$. We show the resulting distributions in the return: $\tilde{\psi}(\lambda/\sigma_{\hat{H}}^2)$, for different values of $\sigma_{\hat{H}}$. In each of the examples in this section $g(x)=1_{x>0}$.
\begin{figure}[H]
\includegraphics[scale=0.65]{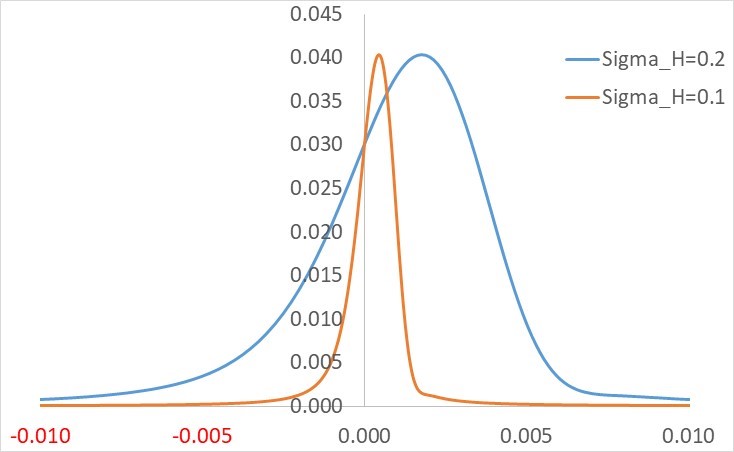}
\caption{The chart shows $\tilde{\psi}(\lambda/\sigma^2)$ for $\sigma_{\hat{H}}=0.2$, and $\sigma_{\hat{H}}=0.1$.}
\end{figure}
The results illustrate that with a non-trivial phase factor, the expected return is no longer zero. Increasing the value of the volatility parameter in the Hamiltonian: $\sigma_{\hat{H}}$ moves the system closer to the near classical limit. The distribution becomes tighter, and the expected value closer to zero.
\vspace{3mm}

The final chart shows the results for $\sigma_{\hat{H}}=0.2$, and the wave function set to: $\psi(x)=(1/\sqrt{2\pi\sigma_S^2})exp((\alpha i-1)x^2/2\sqrt{2}\sigma_S^2)$, with $\alpha=+/-1$. This highlights the fact that the expected value for the return can be positive or negative, depending on the wave function.
\begin{figure}[H]
\centering
\includegraphics[scale=0.65]{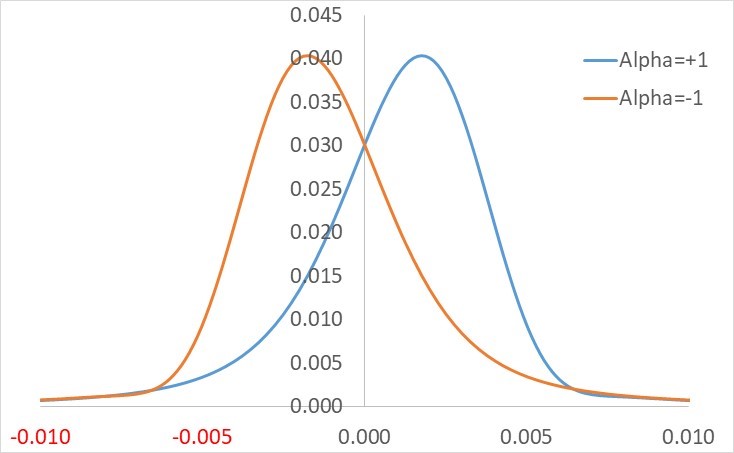}
\caption{The chart shows $\tilde{\psi}(\lambda/\sigma^2)$ for $\alpha=+1$, and $\alpha=-1$.}
\end{figure}
\section{Conclusion}\label{Conc}
In this article we have investigated the randomness that can be introduced through the non-commutativity of the quantum framework, rather than through an external source of noise. We have also shown that, with a real valued market wave function, this randomness will not lead to positive time value in a vanilla option. However, once a non-trivial phase factor: $exp(i\phi(x))$ is applied, the internal randomness can lead to positive or negative time value.

Furthermore, although a real valued wavefunction will mean the expected time value for the option is zero, the spread of potential returns will introduce uncertain time value, once quantum measurement is thrown into the mix.

Going forward, further work is required in a number of areas. For example, this includes, but is not limited to:
\begin{enumerate}
\item[a)] Deciding on how to model trading activity that disturbs the state of the market. For example through frequent measurement of the quantum observables.
\item[b)] Incorporating external environmental noise, for example through quantum stochastic calculus.
\item[c)] The simplest approach may be to combine noise that arises from purely external sources of noise (for example geopolitical events, release of economic data etc) and trading activity, through the methods of quantum stochastic calculus. However, an alternative approach may be to treat these 2 separately.  
\item[d)] Adding more mathematical rigour to the informal geometric approach of section \ref{Cases}.
\end{enumerate}


\end{document}